\documentclass[12pt]{article}
\usepackage{graphicx}
\newcommand{\sinc}{{\rm sinc}}
\newcommand{\ug}{\; = \;}

\newcommand{\Ebf}{\mbox{\boldmath $E$}}

\newcommand{\Hbf}{\mbox{\boldmath $H$}}
\newcommand{\Sbf}{\mbox{\boldmath $S$}}
\newcommand{\Abf}{\mbox{\boldmath $A$}}

\newcommand{\nabf}{\mbox{\boldmath $\nabla$}}

\newcommand{\pa}{\partial}

\newcommand{\text}{\rm}

\newcommand{\drm}{{\rm d}}

\newcommand{\ga}{\gamma}

\newcommand{\la}{\lambda}

\newcommand{\bb}{\begin{equation}}
\newcommand{\ee}{\end{equation}}
\newcommand{\bega}{\begin{eqarray}}
\newcommand{\ega}{\end{eqnarray}}
\newcommand{\begae}{\begin{eqnarray*}}
\newcommand{\egae}{\end{eqnarray*}}

\newcommand{\h}{\hspace*{4ex}}
\newcommand{\dis}{\displaystyle}

\newcommand{\om}{\omega}

\newcommand{\cent}{\centerline}
\newcommand{\vs}{\vspace*}

\begin{document}

\baselineskip 0.6cm

\begin{center}

{\large {\bf Sub-luminal wave bullets: \ Exact Localized subluminal
Solutions to the Wave Equations}}$^{\: (\dag)}$
\footnotetext{$^{\: (\dag)}$  Work partially supported by
FAPESP and CNPq (Brazil), and by INFN, MIUR (Italy). \ E-mail addresses for contacts: mzamboni@ufabc.edu.br,
mzamboni@dmo.fee.unicamp.br [MZR]; \ recami@mi.infn.it [ER]}

\end{center}

\vs{5mm}

\cent{ Michel Zamboni-Rached, }

\vs{2mm}

\centerline{{\em Universidade Federal do ABC, Centro de Sciencias
Naturais e Humanas, Santo Andr\'e, SP;}}
 \cent{{\rm and} DMO--FEEC, State
University at Campinas, Campinas, SP, Brazil.}

\vs{5mm}

\centerline{\rm and}

\vs{3mm}

\cent{ Erasmo Recami }

\vs{2mm}

\cent{{\em Facolt\`a di Ingegneria, Universit\`a statale di Bergamo,
Bergamo, Italy;}}
\cent{{\rm and} {\em INFN---Sezione di Milano, Milan, Italy.}}
%
%
%
%
%
%
%

\vs{8mm}

{\bf Abstract  \ --} \ In this work it is shown how to obtain, in a simple
way, localized (non-diffractive) {\em subluminal} pulses as exact analytic
solutions to the wave equations. These new ideal subluminal solutions,
which propagate without distortion in any homogeneous
linear media, are herein obtained for arbitrarily chosen frequencies
and bandwidths, avoiding in particular any recourse to the non-causal
components so frequently plaguing the previously known localized waves.
The new solutions are suitable superpositions of ---zeroth-order, in general---
Bessel beams, which can be performed either by integrating
with respect to (w.r.t.) the angular frequency $\om$,
or by integrating w.r.t. the longitudinal wavenumber $k_z$: Both methods
are expounded in this paper. \ The first one appears to be powerful
enough; \ we study the second method as well, however, since it allows dealing
even with the limiting case of zero-speed solutions (and furnishes a new way,
in terms of continuous spectra, for obtaining the so-called ``Frozen Waves",
so promising also from the point of view of applications). \  We briefly treat
the case, moreover, of non axially-symmetric solutions, in terms of higher
order Bessel beams. \ At last, particular attention is paid to the role
of Special Relativity, and to the fact that the localized waves are expected to be
transformed one into the other by suitable Lorentz Transformations. \ The analogous
pulses with intrinsic finite energy, or merely truncated, will be constructed in
another paper. \ In this work we fix our attention especially on electromagnetism
and optics: but results of the present kind are valid whenever an
essential role is played by a wave-equation (like in acoustics, seismology,
geophysics, gravitation, elementary particle physics, etc.).\\



\

\

\section{Introduction}

Since more than ten years, the so-called (non-diffracting)
``Localized Waves" (LW), which are new solutions to the
wave equations (scalar, vectorial, spinorial,...), are in fashion,
both in theory and in experiment.  In particular, rather well-known are
the ones with luminal or superluminal peak-velocity: Like the so-called
X-shaped waves (see\cite{Lu1,MRH} and refs. therein; for a review,
see, e.g., ref.\cite{IEEE}), which are supersonic in Acoustics\cite{Lu2},
and superluminal in Electromagnetism (see\cite{PhysicaA} and refs. therein).

\h Since Bateman\cite{Bateman}
and later on Courant \& Hilbert\cite{CH}, it was already known that luminal
LWs exist, which are as well solutions to the wave equations. More
recently, some attention\cite{Mackinnon,deBroglie,Lu5,SaloSalomaa,Longhisub} started
to be paid to {\em subluminal} solutions too. Let us recall that all the
LWs propagate without distortion ---and in a self-reconstructive
way\cite{Bouchal,Grunwald,OpEx2}--- in a homogeneous linear medium
(apart from local variations).

\h Like in the superluminal case, the (more orthodox, in a sense)
subluminal LWs can be obtained by suitable superpositions of
Bessel beams. They have been till now almost neglected, however,
for the mathematical difficulties met in getting analytic
expressions for them, difficulties associated with the fact that
the superposition integral runs over a finite interval. \ We want
here to re-address the question of such subluminal LWs, showing,
by contrast, that one can indeed arrive at exact (analytic)
solutions, both in the case of integration over the Bessel beams'
angular frequency $\om$, and in the case of integration over their
longitudinal wavenumber $k_z$.  The first approach, herein
investigated in detail, is enough to get the majority of the
desired results; we study also the second one, however, since it
allows treating the limiting case of zero-speed solutions (and
furnishes a new method ---based on a continuous spectrum--- for
obtaining the so-called ``Frozen Waves", so promising also from
the point of view of applications). Moreover, we shall deal with
non axially-symmetric solutions, in terms of higher order Bessel
beams. \ At last, particular attention is paid to the role of
Special Relativity, and to the fact that the localized waves are
transformed one into the other by Lorentz Transformations.

\ The present article is devoted to the exact, analytic
solutions: i.e., to ideal solutions. \ In another paper, we shall go on to
the corresponding pulses with finite energy, or truncated, sometimes having
recourse ---for simplicity, and only in such more realistic cases--- to
approximations. \ We shall fix our attention especially on electromagnetism
and optics: but our results are valid whenever an essential role is played
by a wave-equation (like in acoustics, seismology, geophysics, gravitation,
elementary particle physics, etc.).

\h  Let us recall that, in the past, too much attention had not been paid
even to 1983 Brittingham's paper\cite{Brittingham}, where he claimed
the possibility of obtaining pulse-type solutions to the Maxwell equations,
which propagate in free space as a
new kind of speed-c ``solitons".  This was partially due to the fact that that
author was unable to get correct finite-energy expressions for such wavelets,
and to make suggestions about their practical production.  Two years later,
however, Sezginer\cite{Sezginer} was able to show how to obtain quasi-nondiffracting
luminal pulses endowed with a finite energy.
Finite-energy pulses do no longer travel undistorted for an infinite distance,  but they
can nevertheless propagate without deformation for a long field-depth,
much larger than the one achieved by ordinary pulses, as the gaussian ones:
Cf., e.g., refs.\cite{Durnin1,Durnin2,Tesi,Mathieu,Introd,PIER98} and refs.
therein.

\h Only after 1985 the general theory of all LWs started to be extensively
developed\cite{Ziolk1,Ziolk2,Ziolk3,Lu1,Donnelly,PhysicaA,MRH,Esposito,SB1,Friberg}
both in the case of beams, and in the case of pulses. For reviews, see for
instance the refs.\cite{IEEE,Birds,Introd,Tesi,PIER98} and references therein.
 \ For the propagation of LWs in bounded regions (like guides), see
refs.\cite{Coaxial,MRF,MFR,Pahure} and refs therein. \ For the focusing of LWs,
see refs.\cite{MSR,focusing} and refs therein.
As to the construction of LWs propagating in dispersive media, cf.
refs.\cite{Saaridisp,Meiodisp,Porrasdisp,Chirp,Continonlin,Longhidisp,ContiPorras,
FribergPRLnoniso,Capitulo}; and, for lossy media, see ref.\cite{OpEx2} and
refs. therein. \ Al last, for finite-energy, or truncated, solutions
see refs.\cite{ZBS,PIER98,MRH,MRF,OpEx1,JosaMlast}.

\h By now,
the LWs have been experimentally produced\cite{Lu2,PSaari,Ranfagni}, and are
being applied in fields ranging from ultrasound scanning\cite{Lu3,Lu4,Lu5} to
optics (for the production, e.g., of new type of tweezers\cite{brevetto}).
All these works demonstrated that nondiffracting pulses can travel with an
arbitrary peak-speed $v$, that is, with $0<v<\infty$; while Brittingham e
Sezginer had confined themselves to the luminal case ($v=c$) only.
As already commented, the superluminal and luminal LWs have been, and are
being, intensively studied, whilst the subluminal ones have been neglected:
Almost all the few papers dealing with them had till now recourse to the
paraxial\cite{Molone} approximation\cite{paraxial}, or to numerical
simulations\cite{SaloSalomaa}, due to the
above mentioned mathematical difficulty in obtaining exact analytic
expressions for subluminal pulses.  Actually, only {\em one} analytic
solution is known\cite{Mackinnon,Donnelly,deBroglie,Lu5,paraxial} biased by
the physically unconvenient facts that its frequency
spectrum is very large, it doesn't even possess a well-defined central
frequency, and, even more, that non-causal (i.e., backwards
traveling)\cite{Ziolk2,PIER98}
components are a priori needed for constructing it. \ Aim of the present
paper is showing how subluminal localized exact solutions can be constructed
with any spectra,  in any frequency bands and for any bandwidths; and,
moreover, {\em without} employing\cite{MRH,Introd} any non-causal components.

\section{A first method for constructing physically \\
acceptable, subluminal Localized Pulses}

Axially symmetric solutions to the scalar wave equation are known
to be superpositions of zero-order Bessel beams over the angular
frequency $\om$ and the longitudinal wavenumber $k_z$: i.e., in
cylindrical co-ordinates,

\bb \Psi(\rho,z,t) \ug
\int_{0}^{\infty}\,\drm\om\,\int_{-\om/c}^{\om/c}\,\drm k_z\,
\overline{S}(\om,k_z) J_0\left(\rho\sqrt{\frac{\om^2}{c^2} -
k_z^2}\right)e^{ik_z z}e^{-i\om t} \; , \label{eq.(1)} \ee    

where \ $k_\rho^2 \equiv \om^2/c^2 - k_z^2$ \ is the transverse
wavenumber; quantity $k_\rho^2$ has to be positive if one does not want to
deal with evanescent waves.
%

\begin{figure}[!h]
\begin{center}
 \scalebox{.25}{\includegraphics{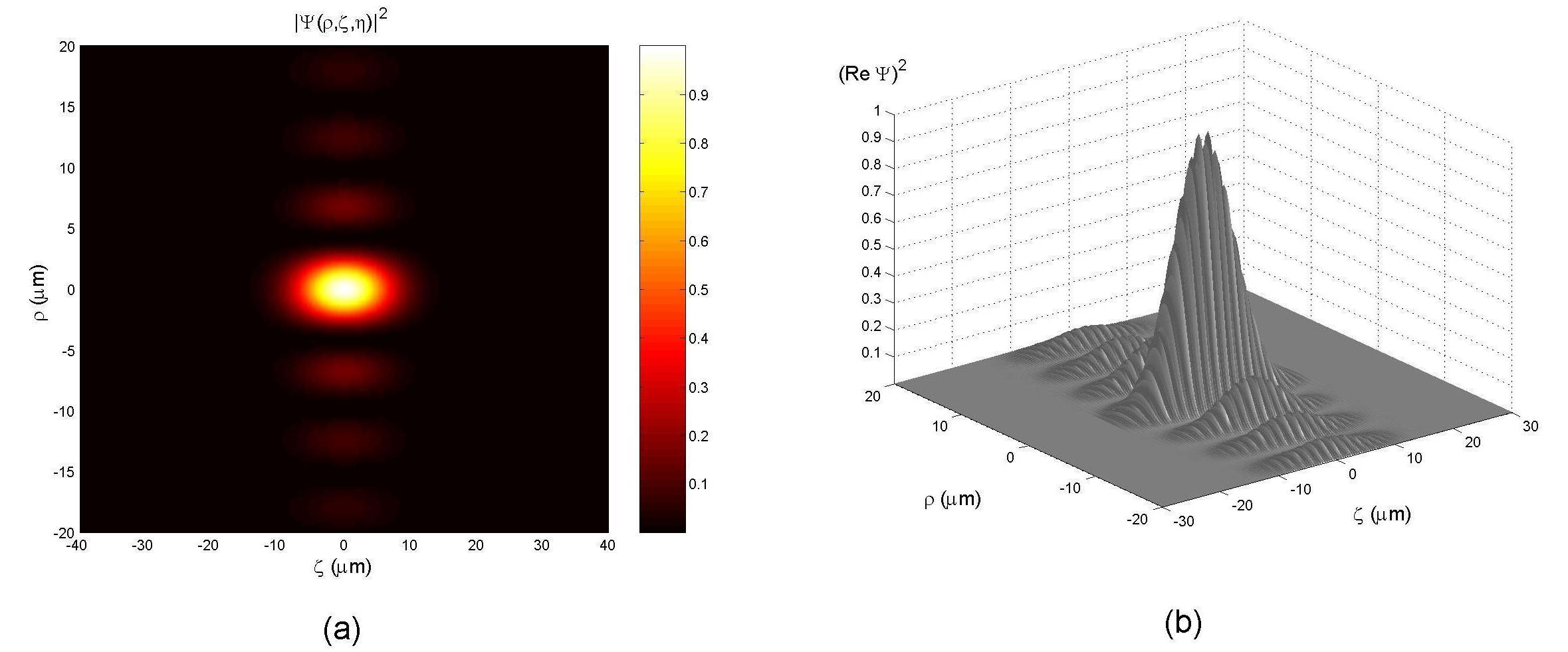}}
\end{center}
\caption{\textbf{(a)} The orthogonal projection of the
pulse intensity; \ \textbf{(b)} The three-dimensional intensity
pattern of the \emph{real part} of the pulse, which reveals the
carrier wave oscillations.} \label{fig1}
\end{figure}

\h The condition characterizing a nondiffracting wave is the
existence\cite{PIER98,MHoptfibers} of a linear
relation between longitudinal wavenumber $k_z$ and frequency $\om$ for
all the Bessel beams entering superposition (1); that is to say, the chosen
spectrum has to entail\cite{MRH,Tesi} for each Bessel beam a linear relationship
of the type\footnote{More generally, as shown in ref.\cite{MRH}, the chosen
spectrum has to call into the play, in the plane $\om, k_z$, if
not exactly the line (2), at least a region in the proximity of a
straight-line of such a type. In the latter case, the corresponding
solution will possess finite energy, but will have a finite ``depth of field":
that is, it will be nondiffracting only till a certain finite distance.} \
(cf. Fig.\ref{fig1}):

\bb \om \ug v\,k_z + b \; \, \label{eq.(2)} \ee      

with $b \geq 0$. \ Requirement (2) can be regarded also as a
specific space-time coupling, implied by the chosen spectrum
$\overline{S}$. Equation (2) has to be obeyed by the spectra of
any one of the three possible types (subluminal, luminal or
superluminal) of nondiffracting pulses. \ Let us mention,
incidentally, that with the choice in Eq.(2) the pulse re-gains
its initial shape after the space-interval ${\Delta z}_1 = 2\pi
v/b$; \ the more general case can be, however,
considered\cite{MRH,Capitulo} when $b$ assumes any values
$b_m=m\,b$ (with $m$ an integer), and the periodicity
space-interval becomes ${\Delta z}_m = {\Delta z}_1 / m\,$.

\h In the subluminal case, of interest here, the only exact solution
known till now, represented by Eq.(10) below, is the one found by
Mackinnon\cite{Mackinnon}. Indeed, by taking into explicit account that the
transverse wavenumber $k_\rho$ of each Bessel beam entering
Eq.(1) has to be real, it can be easily shown (as first noticed by
Salo et al. for the analogous acoustic solutions\cite{SaloSalomaa})
that in the subluminal case $b$ cannot vanish, but must be larger than zero:
 \ $b>0$. \ Then, by using conditions (2) and $b>0$, the subluminal
localized pulses can be expressed as integrals over the frequency only:

\bb \Psi(\rho,z,t) \ug  \exp{[-ib {z \over v}]} \,
\int_{\om_-}^{\om_+}\,\drm\om \; S(\om) \, J_0(\rho k_\rho)\,
\exp{[i\om {\zeta \over v}]} \; , \label{eq.(3)} \ee     

where now

\bb k_\rho \ug {1 \over v} \, \sqrt{2b\om - b^2
- (1-v^2/c^2) \om^2} \; \, \label{eq.(4)} \ee    

with

\bb \zeta \, \equiv \,  z - v t \; \, \label{eq.(5)} \ee    

and with

\hfill{$
\left\{ \begin{array}{clr}
\om_{-} \ug \dis{{b \over {1+v/c}}} \\
\\
\om_{+} \ug \dis{{b \over {1-v/c}}}
\end{array}   \right.
$\hfill} (6)        

\setcounter{equation}{6}

\

As anticipated, the Bessel beam superposition in the subluminal
case results to be an integration over a finite interval of $\om$,
which, by the way, does clearly shows that the non-causal
(backwards traveling) components correspond to the interval
$\om_- < \om < b$. \ It could be noticed, incidentally, that
Eq.(3) does not represent the most general exact solution, which
on the contrary is a sum\cite{Capitulo} of such solutions for the
various possible values of $b$ just mentioned above: That is, for
the values $b_m=m\,b$ and spatial periodicity ${\Delta z}_m =
{\Delta z}_1 / m\,$; but we can confine ourselves to solution (3)
without any real loss of generality, since the actual problem is
evaluating in analytic form the integral entering Eq.(3). For any
mathematical and physical details, see ref.\cite{Capitulo}.

\h Now, if one adopts the change of variables

\bb \om \, \equiv \, {b \over {1-v^2/c^2}}\;(1 + {v \over c} s) \; \,
\label{eq.(7)} \ee    

equation (3) becomes\cite{SaloSalomaa}

\begin{eqnarray}
\lefteqn{\Psi(\rho,z,t) \ug {b \over c} \,{v \over {1-v^2/c^2}} \, \exp{[-i
{b \over v} z]} \;
\exp{\left[ i{b \over v} \, {1 \over {1-v^2/c^2}} \, \zeta \right]}} \nonumber
\\
& & {} \times \int_{-1}^{1}\,\drm s \; S(s) \, J_0\left( {b \over
c} \, {\rho \over {\sqrt{1-v^2/c^2}}} {\sqrt{1-s^2}} \right) \;
\exp{\left[ i {b \over c} {1 \over {1-v^2/c^2}} \zeta s \right]}
\; . \label{eq.(8)}
\end{eqnarray}   

In the following we shall adhere to some symbols standard in
Special Relativity (since the whole topic of subluminal, luminal
and superluminal LWs is strictly connected\cite{IEEE,PhysicaA,RMDartora}
with the principles and structure of
special relativity [cf.\cite{Barut,Review} and refs. therein],
as we shall mention in the Conclusions); namely:

\bb \beta \equiv {v \over c} \; ; \ \ \ \ \ \ \ \ \ \gamma \,
\equiv \, {1 \over {\sqrt{1-\beta^2}}} \; . \label{eq.(9)}
\ee    

\h As already said, Eq.(8) has till now yielded {\em one} analytic solution
for $S(s) \, = \;$constant, only (for instance, $S(s)=1$); which means nothing
but $S(\om) \, = \;$constant: in this case one
gets indeed the Mackinnon solution\cite{Mackinnon,Donnelly,Lu5,JosaMlast}

\begin{eqnarray}
\lefteqn{ \Psi(\rho,\zeta,\eta) \ug 2{b \over c} v \, \gamma^2\,
{\exp{\left[ i{b \over c} \, \beta \gamma^2 \; \eta \right]}}}
\nonumber
 \\
& & {} \times \sinc \ {\sqrt{ {b^2 \over c^2}\, \gamma^2 \left(
\rho^2 + \gamma^2 \; \zeta^2 \right)
}}  \; , \label{eq.(10)}     
\end{eqnarray}

which however, for its above-mentioned drawbacks, is endowed with
little physical and practical interest. \ In Eq.(9) the $\sinc$
function is defined as

\

$$\sinc \ x \, \equiv \, (\sin \; x)/x \; ,$$

\

and

\bb \eta \, \equiv \, z -Vt, \ \ \ \ \ \ \ {\rm with} \ \ V \equiv
{c^2 \over v}
 \; , \label{eq.(11)} \ee  \  
%
%

where $V$ and $v$ are related by the de Broglie relation. \ [Notice that
$\Psi$ in Eq.(10), and in the following ones, is eventually a function
(besides of $\rho$) of $z,t$ via $\zeta$ and $\eta$, both functions of
$z$ and $t\;$].

\h We can, however, construct by a very simple method new
subluminal pulses corresponding to whatever spectrum, and devoid
of non-causal (entering) components, by taking also advantage of
the fact that in our equation (8) the integration interval is
finite: that it, by transforming it in a good, instead of a harm.
Let us first observe that Eq.(10) doesn't admit only the already
known analytic solution corresponding to $S=constant$, and more in
general to $S(\om) \, = \;$constant, but it will also yield an
exact, analytic solution for any exponential spectra of the type

\bb
S(\om) \ug \exp{[{i2n\pi \om \over \Omega}]}
 \; , \label{eq.(12)}
\ee  

with $n$ any integer number, which means for any spectra of the type
$S(s)= \exp{[in\pi / \beta]} \,
\exp{[in\pi s]}$, as can be easily seen by checking the product of the various
exponentials entering the integrand. \ In Eq.(12) we have set

\

$$ \Omega \, \equiv \, {\om_+} - {\om_-} \; . $$

\

The solution writes in this more general case:

\begin{eqnarray}
\lefteqn{ \Psi(\rho,\zeta,\eta) \ug 2b\beta\, \gamma^2 \,
\exp{\left[ i{b \over c} \, \beta \, \gamma^2 \, \eta \right]}}
\nonumber
 \\
& & {} \times \exp{[in {\pi \over \beta}]}
 \ \sinc \ {\sqrt{{b^2 \over c^2}\, \gamma^2 \, \rho^2
+ \left( {b \over c} \, \gamma^2 \, \zeta + n \pi
\right)^2}}
 \; . \label{eq.(13)}
\end{eqnarray}     

Let us explicitly notice that also in Eq.(13) quantity $\eta$ is
defined as in Eqs.(11) above, where $V$ and $v$ obey the de
Broglie relation $vV=c^2$, the subluminal quantity $v$ being the
velocity of the pulse envelope, and $V$ playing the role (in the
envelope's interior) of a superluminal phase velocity.

\h The next step, as anticipated, consists just in taking {\em advantage}
of the finiteness of the integration limits for
expanding any arbitrary spectra $S(\om)$ in a Fourier series in the interval
$\om_{-} \leq \om \leq \om_{+}\;$:

\bb S(\om) \ug \sum_{n=-\infty}^{\infty} \, A_n \,
\exp{[+in {2\pi \over \Omega} \om]} \; , \label{eq.(14)}
\ee    

where (we can go back, now, from the $s$ to the $\om$ variable):

\bb A_n \ug {1 \over \Omega} \, \int_{\om_-}^{\om_+} \drm \om \, S(\om) \,
\exp{[-in {2 \pi \over \Omega} \om]}  \; \,
\label{eq.(15)} \ee    

quantity $\Omega$ being defined as above.

\h Then, on remembering the special solution (13), we can infer from expansion
(14) that, for any arbitrary spectral function $S(\om)$, a rather general,
axially-symmetric, analytic solution for the subluminal case can
be written

\begin{eqnarray}
\lefteqn{ \Psi(\rho,\zeta,\eta) \ug 2b\beta\, \gamma^2 \,
{\exp{\left[ i{b \over c} \, \beta \, \gamma^2 \; \eta \right]}}}
\nonumber
 \\
& & {} \times \sum_{n=-\infty}^{\infty} \, A_n \, \exp{[in{\pi
\over \beta}]} \ \sinc \ {\sqrt{ {b^2 \over c^2}\, \gamma^2 \rho^2
+ \left( {b \over c} \gamma^2 \; \zeta + n \pi \right)^2}}
\; , \label{eq.(16)}     
\end{eqnarray}

in which the coefficients $A_n$ are still given by Eq.(15). Let us
repeat that our solution is expressed in terms of
the particular equation (13), which is a Mackinnon-type solution.
%

\h The present approach presents many advantages. We can easily choose
spectra localized within the prefixed frequency interval (optical waves, microwaves,
etc.) and endowed with the desired bandwidth.   Moreover, as already said,
spectra can now be chosen
such that they have zero value in the region $\om_{-} \leq \om \leq b$,
which is responsible for the non-causal components of the subluminal pulse.

\h Let us stress that, even when the adopted spectrum
$S(\om)$ does not possess a known Fourier series (so that the
coefficients $A_n$ cannot be exactly evaluated via Eq.(15)), one
can calculate approximately such coefficients without meeting any
problem, since our general solutions (8) will still be exact
solutions.

\h Let us set forth in the following some examples.

\

\subsection{Some Examples}

\h In general, optical pulses generated in the lab possess a
spectrum centered on some frequency value, $\om_0$, called the carrier
frequency. \ The pulses can be, for instance, ultra short, when
$\Delta\om/\om_0 \geq 1$; or quasi monochromatic, when
$\Delta\om/\om_0 << 1$, where $\Delta\om$ is the spectrum bandwidth.

\h These kinds of spectra can be mathematically represented by a
gaussian function, or functions with similar behavior.

\

\h {\bf First Example:}

\

Let us consider a gaussian spectrum

\bb S(\om) \ug \frac{a}{\sqrt{\pi}}\exp{[-a^2(\om-\om_0)^2]} \label{eq.(17)} \ee   

whose values are negligible outside the frequency interval $\om_-
< \om < \om_+$ over which the Bessel beams superposition in Eq.(3)
is made, it being  $\om_- = b/(1+\beta)$ and $\om_+ =
b/(1-\beta)$. \ Of course, relation (2) has still to be satisfied,
and with $b>0$, for getting an ideal subluminal localized
solution. \ Notice that, with spectrum (17), the bandwidth
(actually, the FWHM) results to be $\Delta \om = 2 / a$. \ Let us
emphasize that, once $v$ and $b$ have been fixed, the values of
$a$ and $\om_0$ can then be selected in order to kill the
backwards-traveling components, that exist, as we know (cf.
Fig.1) for $\om < b\;$.

\h The Fourier expansion in Eq.(14), which yields, with the above
spectral function (17), the coefficients

\bb A_n \, \simeq \, \dis{{{1} \over {W}}} \, \exp{[-in {2\pi
\over \Omega} \om_0]} \;
\exp{[{{-n^2 \pi^2} \over {a^2 W^2}}]} \; , \label{eq.(18)} \ee   

\

constitutes an excellent representation of the gaussian spectrum (17) in
the interval $\om_- < \om < \om_+$ (provided that, as we requested, our
gaussian spectrum does get negligible values outside the frequency interval
$\om_- < \om < \om_+$).

\h In other words, on choosing a pulse velocity $v<c$ and a value
for the parameter $b$, a subluminal pulse with the above frequency
spectrum (17) can be written as Eq.(15), with the coefficients
$A_n$ given by Eq.(18): The evaluation of such coefficients $A_n$
being rather simple. \ Let us repeat that, even if the values of
the $A_n$ are obtained via a (rather good, by the way)
approximation, we based ourselves on the {\em exact} solution
Eq.(16).

\h One can, for instance, obtain exact solutions representing
subluminal pulses for optical frequencies. Let us get the
subluminal pulse with velocity $v=0.99\;c$, angular carrier
frequency $\om_0= 2.4\times 10^{15}\;$Hz (that is,
$\la_0=0.785\;\mu$m) and bandwidth (FWHM) $\Delta\om=\om_0
/20=1.2\times 10^{14}\;$Hz, which is an optical pulse of $24$ fs
(that is the FWHM of the pulse intensity). \ For a complete pulse
characterization, one has to choose the value of the frequency
$b$: let it be $b=3\times 10^{13}\;$Hz; as a consequence one has
$\om_{-}=1.507\times 10^{13}\;$Hz and $\om_{+}=3\times
10^{15}\;$Hz. \ [This is exactly a case in which the considered
pulse is not plagued by the presence of non-causal components,
since the chosen spectrum forwards totally negligible values for
$\om < b$]. \ The construction of the pulse does already result
satisfactory when considering about 51 terms ($-25 \leq n \leq
25$) in the series entering Eq.(16).

\begin{figure}[!h]
\begin{center}
 \scalebox{.25}{\includegraphics{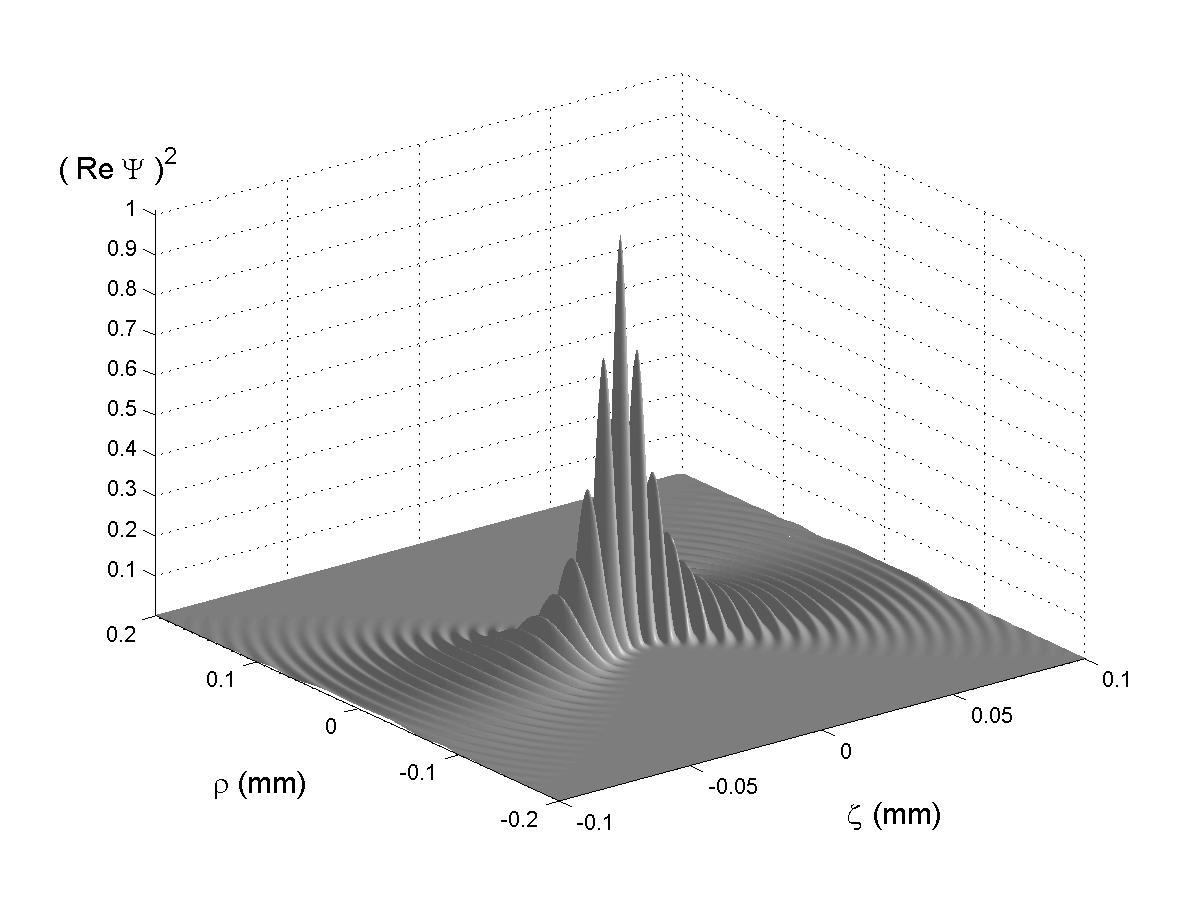}}
\end{center}
\caption{The intensity of the real part of the subluminal pulse
corresponding to spectrum (20), with $v=0.99\;c$,
$b=3\times10^{13}\,$Hz (which result in $\om_-=1.5\times 10^{13}\;$Hz
and $\om_-=3\times 10^{15}\;$Hz), $\Delta\om/\om_+ = 1/100$ (i.e.,
$a=100$).} \label{fig2}
\end{figure}

\h Figures \ref{fig2} show our pulse, plotted by considering the mentioned
fifty-one terms. Namely: Fig.(a) depicts the orthogonal projection
of the pulse intensity; \ Fig.(b) shows the three-dimensional
intensity pattern of the \emph{real part} of the pulse, which
reveals the carrier wave oscillations.

\

\h Let us stress that the ball-like shape\footnote{It can be noted
that each term of the series in Eq.(16) corresponds to an
ellipsoid or, more specifically, to a spheroid, for each velocity
$v$.} for the field intensity should be typically associated with
all the subluminal LWs, while the typical superluminal ones are
known to be X-shaped\cite{Lu1,PhysicaA,RMDartora}, as predicted,
since long, by special relativity in its ``non-restricted"
version: See refs.\cite{Barut,Review,PhysicaA,IEEE} and refs
therein.

\

Another interesting spectrum $S(\om)$ would be, for example, the ``inverted
parabola" one, centered at the frequency $\om_0$: that is,

\bb
S(\om) \ug \left\{ \begin{array}{clr} {\frac{-4 \, [\om -
(\om_0 - \Delta \om/2)] [\om - (\om_0 + \Delta \om/2)]} {\Delta
\om^2}} \;\; & {\rm for}\;\;\; \om_0 - \Delta \om/2 \leq \om \leq
\om_0 + \Delta \om/2 \\
\\
\;\;\;\;\;\; 0  & {\rm otherwise} \ , \label{eq.(19)}
\end{array} \right.    
\ee

where $\Delta \om$, the distance between the two zeros of the
parabola, can be regarded as the spectrum bandwidth. \ One can expand
$S(\om)$, given in Eq.(19), in the
Fourier series (14), for $\om_{-} \leq \om \leq \om_{+}\;$, with coefficients $A_n$
that ---even if straightforwardly calculable--- results to be complicated,
so that we skip reporting them here explicitly. \ Let us here only mention
that spectrum (19) may be easily used to get, for instance, an ultrashort
(femtasecond) optical non-diffracting pulse, with satisfactory results even
when considering very few terms in expansion (14).
%

\

\h {\bf Second Example:}

\

As a second example, let us consider the very simple case when ---within
the integration limits $\om_{-}$, $\om_{+}$---
the complex exponential spectrum (12) is replaced by the real function
(still linear in $\om$)

\bb S(\om) \ug \frac{a}{1-\exp{[-a(\om_+ - \om_-)]}}\,\exp{[a(\om
- \om_{+}]}
 \; , \label{eq.(20)}
\ee  

with $a$ a positive number [for $a=0$ one goes back to the
Mackinnon case]. Spectrum (20) is exponentially concentrated in
the proximity of $\om_{+}$, where it reaches its maximum value;
and becomes more and more concentrated (on the left of $\om_{+}$,
of course) as the arbitrarily chosen value of $a$ increases, with
the frequency bandwidth being $\Delta\om=1/a$. Let us recall,
incidentally, that, on its turn, quantity $\om_{+}$ and $\om_{-}$
depend on the pulse velocity $v$ and on the arbitrary parameter
$b$.

\h By performing the integration as in the case of spectrum (12), instead
of solution (13) in the present case one eventually gets the solution

\begin{eqnarray}
\lefteqn{ \Psi(\rho,\zeta,\eta) \ug \frac{2ab\beta\gamma^2 \,
\exp{[ab\gamma^2]} \, \exp{[-a\om_{+}]}}{1-\exp{[-a(\om_+ -
\om_-)]}}}
 \nonumber
 \\
 \nonumber
 \\
 & & {} \times
 \exp{\left[ i{b \over c} \, \beta \, \gamma^2 \, \eta \right]}
 \ \sinc{\left[ {b \over c} \, \gamma^2 \; {\sqrt{\gamma^{-2} \, \rho^2
- (av+i\zeta)^2}} \right]} \; . \label{eq.(21)}
\end{eqnarray}     

\h After Mckinnon's, this eq.(21) appears to be the simplest
closed-form solution, since both of them do not need any recourse
to series expansions. In a sense, our solution (21) might be
regarded as the subluminal analogous of the (superluminal) X-wave
solution; a difference being that the standard X-shaped solution
has a spectrum starting with 0, where it assumes its maximum
value, while in the present case the spectrum starts at $\om_{-}$
and gets increasing afterwards till $\om_+$  . More important is
to observe that the gaussian spectrum has a priori two advantages
w.r.t. eq.(20): It may be more easily centered around any value
$\om_{0}$ of $\om$, and, when increasing its concentration in the
surrounding of $\om_{0}$, the spot transverse width does not
increase indefinitely, but tends to the spot width of a Bessel
beam with $\om=\om_0$ and $k_z=(\om_0 - b)/V$, at variance with
what happens for spectrum (20); however, solution (21) is
noticeable, since it is really the simplest one.

\begin{figure}[!h]
\begin{center}
 \scalebox{.25}{\includegraphics{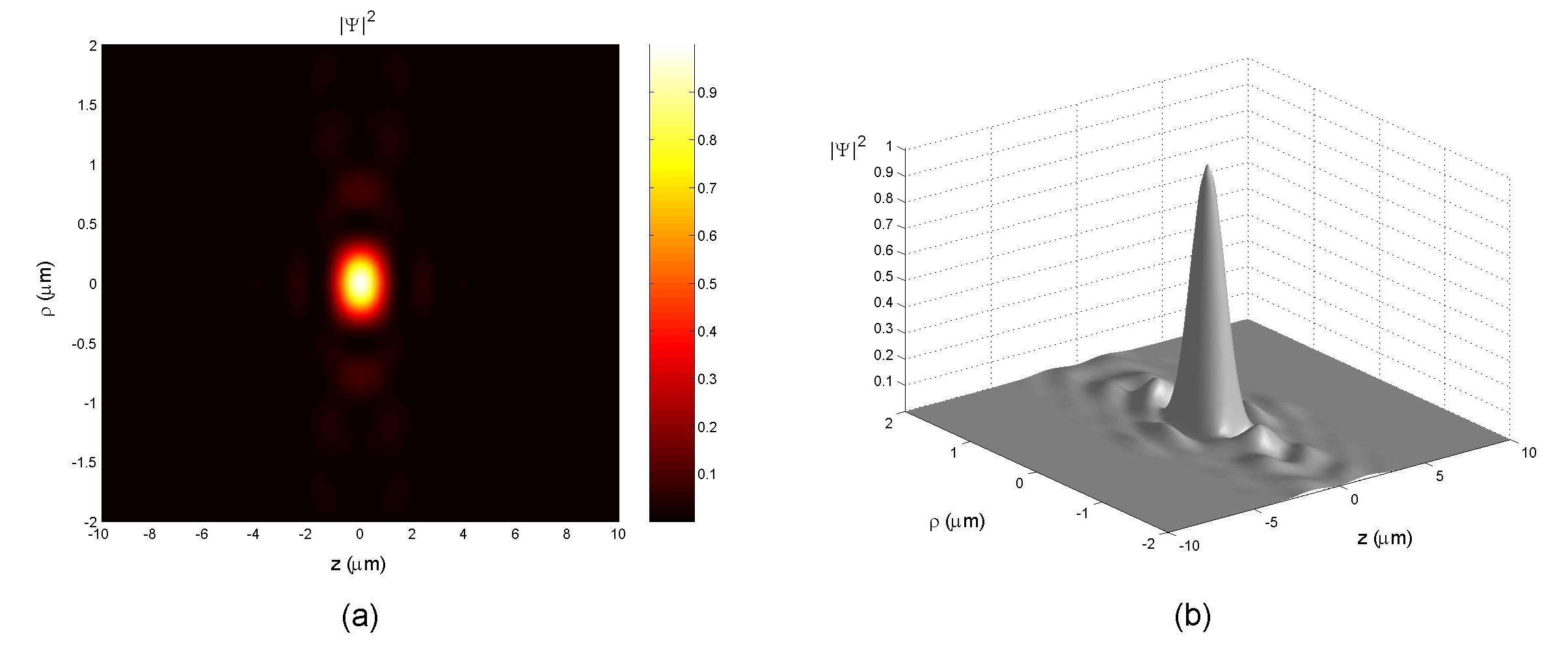}}
\end{center}
\caption{\textbf{(a)} Orthogonal projection of the
three-dimensional intensity pattern of the beam (a null-speed
subluminal wave) corresponding to spectrum (32); \ \textbf{(b)} 3D
plot of the field intensity. The beam considered in this example
is highly non-paraxial.} \label{fig3}
\end{figure}

\h Figure \ref{fig3} show the intensity of the real part of the subluminal
pulse corresponding to this spectrum, with $v=0.99\;c$,
$b=3\times10^{13}\;$Hz (which result in $\om_-=1.5\times 10^{13}\;$Hz
and $\om_-=3\times 10^{15}\;$Hz), $\Delta\om/\om_+ = 1/100$ (i.e.,
$a=100$). This is an optical pulse of $0.2$ ps.

\

\section{A second method for constructing\\
subluminal Localized Pulses}

The previous method appears to be very efficient for finding out analytic
subluminal LWs, but it looses its validity in the limiting case $v
\rightarrow 0$, since for $v=0$ it is $\om_{-} \equiv \om_{+}$ and the
integral in Eq.(3) degenerates, furnishing a null value. \ By
contrast, we are interested also in the $v=0$ case, since it corresponds
to some of the most interesting, and potentially useful, LWs: Namely,
to the so-called ``Frozen Waves", which are new stationary solutions
to the wave equations, possessing a static envelope.

\h It is possible to solve this problem by starting again from Eq.(1),
with constraint (2), but going on ---this time--- to integrals over
$k_z$. instead of over $\om$. \ It is enough to write relation (2)
in the form

\hfill{$
k_z \ug \dis{{1 \over v} \, (\om - b)}
$\hfill} (2')        

for expressing the exact solutions (1) as

\bb \Psi(\rho,z,t) \ug  \exp{[-ibt]} \;
\int_{k_z{\rm min}}^{k_z{\rm max}}\,\drm k_z \; S(k_z) \, J_0(\rho k_\rho)\,
\exp{[i \zeta k_z]} \; , \label{eq.(22)} \ee     

with

\hfill{$
\left\{ \begin{array}{clr}
k_z{\rm min} \ug \dis{ {-b \over c} \, {1 \over {1+\beta}}} \\
\\
k_z{\rm max} \ug \dis{ {b \over c} \, {1 \over {1-\beta}} }
\end{array}   \right.
$\hfill} (23)        

\setcounter{equation}{23}

and with

\bb {k_\rho}^2 \ug -{k_z^2 \over \gamma^2} + 2 {b \over c} \beta k_z +
{b^2 \over c^2} \; , \label{eq.(24)} \ee    

where quantity $\zeta$ is still defined according to Eq.(5), always with
$v<c$.

\h One can show that the unique exact solution previously
known\cite{Mackinnon} may be rewritten in the form of Eq.(24) {\em with}
$S(k_z) = \;$constant. \ Then, on following the same procedure exploited
in our first method (previous Section), one can find out new exact solutions
corresponding to

\bb
S(k_z) \ug \exp{[{i2n\pi k_z \over K}]}
 \; , \label{eq.(25)}
\ee  

where

\

$$ K \, \equiv \, k_z{\rm max} - {k_z{\rm min}} \; , $$

\

by performing the change of variable [analogous, in its finality, to the
one in Eq.(7)]

\bb k_z \, \equiv \, {b \over c} \, \gamma^2 \, (s + \beta) \; .
\label{eq.(26)} \ee    

\h At the end, the exact subluminal solution corresponding to the
spectrum (25) results to be

\begin{eqnarray}
\lefteqn{ \Psi(\rho,\zeta,\eta) \ug 2\,{b \over c} \, \gamma^2 \,
\exp{[i{b \over c} \, \beta \, \gamma^2 \, \eta]}}  \nonumber
 \\
& & {} \times \exp{[in \pi \beta]}
 \ \sinc \ {\sqrt{{b^2 \over c^2}\, \gamma^2 \, \rho^2 +
\left( {b \over c} \, \gamma^2 \, \zeta + n \pi \right)^2}}
 \; , \label{eq.(27)}
\end{eqnarray}          

\h We can again observe that any spectra $S(k_z)$ can be expanded,
in the interval $k_z{\rm min} < k_z < k_z{\rm max}$, in a Fourier
series:

\bb S(k_z) \ug \sum_{n=-\infty}^{\infty} \, A_n \,
\exp{[+in {2\pi \over K} k_z]} \; , \label{eq.(28)}
\ee    

with coefficients given now by

\bb A_n \ug {1 \over K} \, \int_{k_z{\rm min}}^{k_z{\rm max}} \drm k_z \,
S(k_z) \, \exp{[-in {2 \pi \over K} k_z]} \; \,
\label{eq.(29)} \ee    

quantity $K$ having been defined above.

\h At the very end of the whole procedure, the general exact solution
representing a subluminal LW, for any spectra $S(k_z)$, can be
eventually written

\begin{eqnarray}
\lefteqn{ \Psi(\rho,\zeta,\eta) \ug 2\,{b \over c} \, \gamma^2 \,
\exp{[i{b \over c} \, \beta \, \gamma^2 \, \eta]}}  \nonumber
 \\
& & {} \times \sum_{n=-\infty}^{\infty} \, A_n \, \exp{[in \pi \beta]}
 \ \sinc \ {\sqrt{{b^2 \over c^2}\, \gamma^2 \, \rho^2 +
\left( {b \over c} \, \gamma^2 \, \zeta + n \pi \right)^2}}
 \; , \label{eq.(30)}
\end{eqnarray}     

whose coefficients are expressed in Eq.(29), and where quantity $\eta$
is defined as above, in Eq.(11).

\h Interesting examples could be easily worked out, as we did at the end of
the previous Section.

\section{\bf Stationary solutions with zero-speed envelopes\\
(``Frozen Waves")}

Here, we shall refer ourselves to the (second) method, expounded
in the previous Section. \ Our solution (30), for the case of
envelopes {\em at rest}, that is, in the case $v=0$ [which implies
$\zeta = z$], becomes

\

\hfill{$
\Psi(\rho,z,t) \ug \dis{ 2\,{b \over c} \; \exp{[-i b t]} \; \sum_{n=-\infty}^{\infty} \, A_n \
\sinc \ {\sqrt{{b^2 \over c^2}\, \rho^2 +
\left( {b \over c} \, z + n \pi \right)^2}}} \; ,
$\hfill} (31)        

\

with coefficients $A_n$ given by Eq.(29) with $v=0$, so that its integration
limits simplify into $-b/c$ and $b/c$, respectively; thus, one gets

\

\hfill{$ A_n \ug \dis{{c \over {2b}} \, \int_{-b/c}^{b/c} \drm k_z \,
S(k_z) \, \exp{[-in {c \pi \over b} k_z]}} \; .
$\hfill} (29')        

\

\h Equation (31) is a new, exact solution, corresponding to
stationary beams with a {\em static} intensity envelope. \ Let us
observe, however, that even in this case one has energy
propagation, as it can be easily verified from the power flux
$\Sbf = -\nabf\Psi \ \pa\Psi/\pa t$ (scalar case) or from the
Poynting flux $\Sbf = (\Ebf \wedge \Hbf)$ (vectorial case: the
condition being that $\Psi$ be a single component, $A_z$, of the
vector potential $\Abf$).\cite{PhysicaA} \ For $v=0$, eq.(2)
becomes

$$ \om \ug b \, \equiv \, \om_0 \; ,$$

so that the particular subluminal waves endowed with null velocity are
actually monochromatic beams.

\h It may be stressed that the present (second) method does yield
{\em exact} solutions, without any need of the paraxial approximation,
which, on the contrary, is so often used when looking for
expressions representing beams, like the gaussian ones. \ Let us
recall that, when having recourse to the paraxial approximation,
the obtained beam expressions are valid only when the envelope
sizes (e.g., the beam spot) vary in space much more slowly than
the beam wavelength.  For example, the usual expression for a
gaussian beam\cite{Molone} holds only when the beam spot $\Delta
\rho$ is much larger than $\lambda_0 \equiv \om_0 / (2\pi c) =
b/(2\pi c)$: so that those beams {\em cannot} be very much
localized. \ By contrast, our method overcomes such problems,
since it provides us, as we have seen above, with exact
expressions for (well localized) beams with sizes {\em of the
order} of their wavelength. \ Notice, moreover, that the
already-known exact solutions ---for instance, the Bessel beams---
are nothing but particular cases of our solution (31).

\

{\em An example:} \ On choosing (with $0 \leq q_- < q_+ \leq 1$)
the spectral double-step function

\

\hfill{$
S(k_z) \ug \left\{ \begin{array}{clr}
\dis{\frac{c}{\om_0(q_+ - q_-)}} \ \ \ \ \ \ \ \ \ \ & {\rm for} \ q_-\om_0/c \leq k_z \leq q_+\om_0/c \\
\\
0 \ \ \ \ \ \ \ \ \ \ & {\rm elsewhere} \; ,
\end{array}   \right.
$\hfill} (32)        

\

the coefficients of Eq.(31) become

\

\hfill{$ A_n \ug \dis{\frac{ic}{2\pi n\om_0(q_+ - q_-)} \; \left[
e^{-iq_+\pi n} - e^{-iq_-\pi n} \right]}
 \; .
$\hfill} (33)        

\

\h The double-step spectrum (32), with regard to the longitudinal wave number,
corresponds to the mean value $\overline{k}_z = \om_0(q_++q_-)/2c$ and to
the width $\Delta k_z=\om_0(q_+-q_-)/c$. From these relations, it follows
that $\Delta k_z/\overline{k}_z = 2(q_+-q_-)/(q_++q_-)$.

\h For values of $q_-$ and $q_+$ that do not satisfy the inequality $\Delta
k_z / \overline{k}_z << 1$, the resulting solution will be a {\em non-paraxial}
beam.

\h Figures 3 show the exact solution corresponding to $\om_0 =
1.88\times 10^{15}\;$Hz (i.e., $\lambda_0=1\;\mu$m) and to $q_- =
0.3$, $q_+ = 0.9$, which results to be a beam with a spot-size
diameter of $0.6\;\mu$m, and, moreover, with a rather good
longitudinal localization. \ In the case of Eqs.(32,33), about 21
terms ($-10 \leq n \leq 10$) in the sum entering Eq.(31) are quite
enough for a good evaluation of the series. The beam considered in
this example is highly non-paraxial (with $\Delta
k_z/\overline{k}_z = 1$ ), and therefore couldn't have been
obtained by ordinary gaussian beam solutions (which are valid in
the paraxial regime only)\footnote{We are considering here only
scalar wave fields. In the case of non-paraxial optical beams, the
vector character of the field has to be considered}.

\

\h Let us now emphasize that a noticeable property of our present
method is that it allows a spatial modeling even of monochromatic
fields (that correspond to envelopes {\em at rest\/}; so that, in the
electromagnetic cases, one can speak, e.g., of the modeling of
``light-fields at rest"). \ Such a property ---rather interesting,
especially for applications\cite{brevetto}--- was already
investigated, under different assumptions, in
refs.\cite{OpEx1,FWart2,OpEx2}, where the stationary fields with
static envelopes were called ``Frozen Waves" (FW). \ Namely, in
the quoted references, discrete superpositions of Bessel beams
were adopted in order to get a predetermined longitudinal
(on-axis) intensity pattern, inside the desired space interval
$0<z<L$. \ In other words, in refs.\cite{OpEx1,FWart2,OpEx2}, the
Frozen Waves have been written in the form

\

\hfill{$
\begin{array}{clr}
\Psi(\rho,z,t) \ug \dis{ e^{-i\om_0t} \ e^{iQz} }  \\
 \\
\ \ \ \ \ \ \ \ \ \ \times \dis{ \sum_{n=-N}^{N} \, B_n \; J_0(\rho {k_\rho}_n) \;
e^{i2n\pi z/c} }
\end{array}
$\hfill} (34)        

\

with

\

\hfill{$
B_n \ug
\dis{ {1 \over L} \, \int_{0}^{L} \drm z \; F(z) \ e^{-i2n\pi z/L} }  \; ,
$\hfill} (35)    

\

quantity $|F(z)|^2$ being the desired longitudinal intensity
shape, chosen a priori. \ In Eq.(35), it is \ ${{k_\rho}_n}^2 =
n^2 \om^2/c^2 - {{k_z}_n}^2$, \ and \ $0 \leq Q+2N\pi/L \leq
\om_0/c$, \ were we put \ ${{k_z}_n} \equiv Q+2n\pi /n$. \ As we
see from Eq.(34), the FWs have been represented in the past in
terms of {\em discrete} superpositions of Bessel beams. \ But,
now, the method exploited in this paper allows us to go on to
dealing with {\em continuous} superpositions. In fact, the
continuous superposition analogous to eq.(34) would write

\

\hfill{$ \Psi(\rho,z,t) \ug \dis{ e^{-i\om_0 t} \;
\int_{-\om_0/c}^{\om_0/c} \drm k_z \ S(k_z) \ J_0(\rho k_\rho) \
e^{iz k_z} } \; ,
$\hfill} (36)    

\

which, actually, is nothing but our previous Eq.(22) with $v=0$
(and therefore $\zeta = z$): \ That is, eq.(36) does just
represent a {\em null-speed} subluminal wave. \ To be clearer, let
us recall that the FWs were expressed in the past as discrete superposition,
mainly because it was not known at that time how to treat
analytically a continuous superposition like eq.(36). Only by
following the method presented in this work one can eventually
extend the FW approach\cite{OpEx1,FWart2,OpEx2} to the case of
integrals: without numerical simulations, but in terms once more
of analytic solutions.

\h Indeed, the exact solution of Eq.(36) is given by Eq.(31), with
coefficients (29').  And one can choose the spectral function
$S(k_z)$ in such a way that $\Psi$ assumes the on-axis pre-chosen
static intensity pattern $|F(z)|^2$. Namely, the equation to be
satisfied by $S(k_z)$, to such an aim, comes by associating
eq.(36) with the requirement $|\Psi(\rho=0,z,t)|^2 = |F(z)|^2$,
which entails the integral relation

\

\hfill{$  \dis{\int_{-\om_0/c}^{\om_0/c} \drm k_z \; S(k_z) \;
e^{i z k_z}} \ug F(z) \; .
$\hfill} (37)    

\

Equation (37) would be trivially soluble in the case of an
integration between $-\infty$ and $+\infty$, since it would merely
be a Fourier transformation; but obviously this is not the case,
because its integration limits are finite. Actually, there are
functions $F(z)$ for which Eq.(37) is not soluble, in the sense
that no spectra $S(k_z)$ exist obeying the last equation. Namely,
if we consider the {\em Fourier} expansion

\

\hfill{$ F(z) \ug \dis{\int_{-\infty}^{\infty}} \drm k_z \;
\widetilde{S}(k_z) \; e^{iz k_z} \; , $\hfill}

\

when $\widetilde{S}(k_z)$ does assume non-negligible values outside
the interval $-\om_0/c < k_z < \om_0/c$, then in Eq.(37) no $S(k_z)$
can forward that particular $F(z)$ as a result.

\h {\em However}, way-outs can be devised, such that one can can
nevertheless find out a function $S(k_z)$ that approximately
(but satisfactorily) complies with our Eq.(37).

\h The first way-out consists in writing $S(k_z)$ in the form

\

\hfill{$ S(k_z) \ug \dis{ {1 \over K} \; \sum_{n=-\infty}^{\infty}
\, F \left( {{2n\pi} \over K} \right) \; e^{-i2n\pi k_z / K} } \;
.
$\hfill} (38)    

\

where, as before, $K=2\om_0/c$. Then, one can easily verify
Eq.(38) to guarantee that the integral in Eq.(37) yields the
values of the desired $F(z)$ {\em at the discrete points} \ $z =
2n\pi / K$. \ Indeed, the Fourier expansion (38) is already of the
same type as Eq.(28), so that in this case the coefficients $A_n$
of our solution (31), appearing in Eq.(29'), do simply become

\

\hfill{$
A_n \ug \dis{{1 \over K} \; F(-{{2n\pi} \over K})} \; .
$\hfill} (39)    

\

\h This is a powerful way for obtaining a desired longitudinal (on-axis)
intensity pattern, especially for very small spatial
regions, because it is not necessary to solve any integral to find
out the coefficients $A_n$, which by contrast are given directly by Eq.(39).

\begin{figure}[!h]
\begin{center}
 \scalebox{.8}{\includegraphics{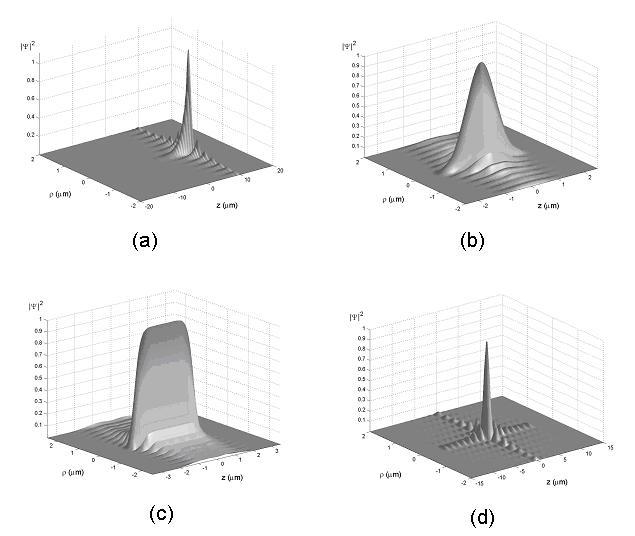}}
\end{center}
\caption{Frozen Waves with the on-axis longitudinal field pattern
chosen as: \textbf{(a)} Exponential; \ \textbf{(b)} Gaussian;  \
\textbf{(c)} Super-gaussian;  \ \textbf{(d)} Zero order \ Bessel
function} \label{fig4}
\end{figure}

\h Figures \ref{fig4} depict some interesting applications of this method.
A few  desired longitudinal intensity patterns $|F(z)|^2$ have been chosen,
and the corresponding Frozen Waves calculated by using Eq.(31) with the
coefficients $A_n$ given in Eq.(39). The desired patterns are enforced
to exist within very small spatial intervals only, in order to show the
capability of our method to model the field intensity shape also under
such strict requirements.

\h In the four examples below, we considered a wavelength
$\lambda=0.6\;\mu$m, which corresponds to $\om_0=b=3.14\times
10^{15}\;$Hz.

\h The first longitudinal (on-axis) pattern considered by us is that given
by

\

$$
 F(z) \ug \left\{\begin{array}{clr}
 \dis{e^{a(z-Z)}} \;\;\; &
 {\rm for}\;\;\; 0 \leq z \leq Z  \\

 \\
 \;\;\;\;\;\;\;\; 0  & \mbox{elsewhere}
\end{array} \right. \; , \label{Fz1}
 $$

i.e., a pattern with an exponential increase, starting from $z=0$
till $z=Z$. The chosen values of $a$ and $Z$ are $Z=10\;\mu$m
and $a=3/Z$. The intensity of the corresponding Frozen Wave is
shown in Fig.(4a).

\h The second longitudinal pattern (on-axis) taken into consideration
is the gaussian one, given by

\

$$
 F(z) \ug \left\{\begin{array}{clr}
\dis{e^{-q(\frac{z}{Z})^2}} \;\;\; &
 {\rm for}\;\;\; -Z \leq z \leq Z  \\

 \\
 \;\;\;\;\;\;\;\; 0  & \mbox{elsewhere}
\end{array} \right. \; , \label{Fz1}
 $$

 \

with $q=2$ and $Z=1.6\;\mu$m. The intensity of the corresponding
Frozen Wave is shown in Fig.(4b).

\h In the third example, the desired longitudinal pattern is supposed to
be a super-gaussian:

\

$$
 F(z) \ug \left\{\begin{array}{clr}
 \dis{e^{-q(\frac{z}{Z})^{2m}}} \;\;\; &
 {\rm for}\;\;\; -Z \leq z \leq Z  \\

 \\
 \;\;\;\;\;\;\;\; 0  & \mbox{elsewhere \ ,}
\end{array} \right. \; , \label{Fz1}
 $$

 \

where $m$ controls the edge sharpness. Here we have chosen
$q=2$, $m=4$ and $Z=2 \ \mu$m. \ The intensity of the Frozen Wave
obtained in this case is shown in Fig.(4c).

\h Finally, in the fourth example, let us choose the longitudinal
pattern as being the zero-order Bessel function

\

$$
 F(z) \ug \left\{\begin{array}{clr}
 J_0(q\,z) \;\;\; &
 {\rm for}\;\;\; -Z \leq z \leq Z  \\

 \\
 \;\;\;\;\;\;\;\; 0  & \mbox{elsewhere}
\end{array} \right. \; , \label{Fz1}
 $$

with $q=1.6\times 10^{6}\;\rm{m}^{-1}$ and $Z=15\;\mu$m. The intensity
of the corresponding Frozen Wave is shown in Fig.(4d).

\h Let us observe that, of course, any static envelopes of this type can
be easily transformed into propagating pulses by the mere application
of Lorentz transformations.

\

\

\h Another way-out exists for evaluating $S(k_z)$, based on the assumption
that

\

\hfill{$
S(k_z) \, \simeq \, \widetilde{S}(k_z)  \; ,
$\hfill} (40)    

\setcounter{equation}{40}

\

which consitutes a good approximation whenever $\widetilde{S}(k_z)$ assumes
{\em negligible} values outside the interval $[-\om_0/c, \ \om_0/c]$. \ In
such a case, one can have recourse to the method associated with Eq.(28)
and expand $\widetilde{S}(k_z)$ itself in a Fourier series, getting
eventually the relevant coefficients $A_n$ by Eq.(29). \ Let us recall
that it is still $K \equiv k_z{\rm max} - k_z{\rm min} = 2\om_0/c$.

\h It may be interesting to call attention to the circumstance
that, when constructing FWs in terms of a sum of discrete
superpositions of Bessel beams (as it was done by us in
refs.\cite{OpEx1,FWart2,OpEx2,brevetto}), it was easy to obtain
extended envelopes like, e.g., ``cigars": where easy means having
recourse to few terms of the sum. By contrast, when we construct
FWs ---following this Section--- as continuous superpositions,
then it is easy to get higly localized (concentrated) envelopes. \
Let us explicitly mention, moreover, that the method presented in
this Section furnishes FWs that are no longer periodic along the
$z$-axis (a situation that, with our old method,\cite{OpEx1,FWart2,OpEx2}
was obtainable only when the periodicity interval tended to infinity).

\

\

\section{The role of Special Relativity, and of Lorentz Transformations}

Strict connections exist between, on one hand, the principles and
structure of Special Relativity and, on the other hand, the whole
subject of subluminal, luminal, superluminal Localized Waves, in
the sense that it is expected since long time that a priori they
are transformable one into the other via suitable Lorentz
transformations (cf. refs.\cite{Barut,Review,NCim}, besides work
of ours in progress).\footnote{After the completion of this work,
we realized that we had overlooked an important and interesting
paper by Saari et al.\cite{Saari2004}, wherein the relativistic
connections between the LWs are already, deeply and in general,
investigated in terms of suitable LTs: we therefore inserted in
this article the necessary quotations. We are actually glad in
calling attention to ref.\cite{Saari2004}, also because the
inspiring philosophy ---which in part goes back to papers like
refs.\cite{Barut,Review}--- is shared by us too.}

\h Let us first confine ourseves to the cases faced in this paper. Our
subluminal localized pulses, that may be called ``wave bullets",
behave as {\em particles\/}: Indeed, our subluminal pulses [as well as the
luminal and superluminal (X-shaped) ones, that have been amply investigated
in the past literature] do exist as solutions of any wave equations, ranging
from electromagnetism and acoustics or geophysics, to elementary particle
physics (and even, as we discovered recently, to gravitation physics). \
From the kinematical point of view, the velocity composition relativistic
law holds also for them. \ The same is true, more in general,
for any localized waves (pulses or beams).

\h Let us start for simplicity by considering, in an initial reference-frame
O, just a ($\nu$-order) Bessel beam:

\bb
\Psi(\rho,\phi,z,t) \ug J_\nu(\rho k_\rho) \; e^{i\nu\phi} \; e^{iz k_z} \;
e^{-i\om t}
\; . \label{eq.(41)} \ee    

In a second reference-frame O', moving with respect to (w.r.t.) O
with speed $u$ ---along the positive z-axis and in the positive
direction, for simplicity's sake---, it will be observed\cite{Saari2004} the new Bessel beam

\bb \Psi(\rho ',\phi ',z ',t ') \ug J_\nu(\rho ' {k'}_{\rho'}) \;
e^{i\nu\phi '} \; e^{iz' {k'}_{z'}} \; e^{-i\om ' t'} \; ,
\label{eq.(42)} \ee    

obtained by applying the appropriate Lorentz transformation (a Lorentz
``boost") with \ $\gamma = [\sqrt{1-u^2/c^2}]^{-1}$:

\bb {k'}_{\rho'} = k_\rho; \ \ \ {k'}_{z'} = \ga (k_z-u\om/c^2); \
\ \ \om ' = \ga (\om - uk_z)  \; ;
\label{eq.(43)} \ee    

this can be easily seen, e.g., by putting

\bb
\rho = \rho '; \ \ \  z = \ga (z'+ut'); \ \ \  t = \ga (t'+ uz'/c^2)
\label{eq.(44)} \ee    

directly into Eq.(42).

\h Let us now pass\cite{Saari2004} to subluminal {\em pulses}. We can investigate
the action of a Lorentz transformation (LT), by expressing them
either via the first method (Section 2) or via the second one
(Section 3). \ Let us consider for instance, in the frame O,
a $v$-speed (subluminal) pulse, given by Eq.(3) of our Section 2. \
When we go on to
a second observer O' moving {\em with the same speed} $v$ w.r.t. frame
O, and, still for the sake of simplicity, passing through the origin $O$ of
the initial frame at time $t=0$, the new observer O' will see
the pulse\cite{Saari2004}

\bb \Psi(\rho ',z ',t ') \ug e^{-i t' {\om'}_0} \,
\int_{\om_{-}}^{\om_{+}} \drm \om \; S(\om) \; J_0(\rho '
{k'}_{\rho'}) \; e^{i z' {k'}_{z'}}
\; , \label{eq.(45)} \ee     

with

\bb
{k'}_{z'} = {\ga}^{-1} \om/v - \ga b/v; \ \ \ \om ' = \ga b; \ \ \
{k'}_{\rho'} = {\om'}_0/c^2 - {{k'}_{z'}}^2 \; ,
\label{eq.(46)} \ee    

as one gets from the Lorentz transformation in Eq.(43), or in Eq.(44),
with $u = v$ \ [and $\gamma$ given by Eqs.(9)]. \ Notice that ${k'}_{z'}$
is a function of $\om$, as expressed by the first one of the three
relations in the previous Eqs.(46); and that here $\om'$ is a constant.

\h If we explicitly insert into Eq.(45) the relation \ $ \om =
\ga (v{{k'}_{z'}} + \ga b)$, \ which is nothing but a re-writing of the first
one of Eqs.(46), then Eq.(45) becomes\cite{Saari2004}

\bb \Psi(\rho ',z ',t ') \ug \ga v \; e^{-i t' \om_0} \;
\int_{-\om_0/c}^{\om_0/c} \drm {k'}_{z'} \;
\overline{S}({k'}_{z'}) \; J_0(\rho ' {k'}_{\rho'}) \; e^{i z'
{k'}_{z'}} \; ,
\label{eq.(47)} \ee     

where $\overline{S}$ is expressed in terms of the previous function $S(\om)$,
entering Eq.(45), as follows:

\bb
\overline{S}({k'}_{z'}) \ug S(\ga v {k'}_{z'} + \ga^2 b)
\; . \label{eq.(48)} \ee     

Equation (47) describes monochromatic beams with axial symmetry
(and does coincide also with what derived within our second
method, in Section 3, when posing $v=0$).

\h The remarkable conclusion is that a subluminal pulse, given by
our Eq.(3), which appears as a $v$-speed {\em pulse} in a frame O,
will appear\cite{Saari2004} in another frame O' (traveling w.r.t.
observer O with the same speed $v$ in the same direction $z$) just
as the {\em monochromatic beam} in Eq.(47) endowed with angular
frequency ${\om'}_0 = \ga b$, whatever be the pulse spectral
function in the initial frame O: even if the kind of monochromatic
beam one arrives to does of course depend\footnote{One gets in
particular a Bessel-type beam when $S$ is a Dirac's
delta-function: $S(\om)= \delta(\om-\om_0)$. Moreover, let us
notice that, on applying a LT to a Bessel beam, one obtains
another Bessel beam, with a different axicon-angle.}  on the
chosen $S(\om)$. The vice-versa is also true, in general.\

\h Let us set forth explicitly an observation that hasn't been
noticed in the existing literature yet. Namely, let us mention
that, when starting not from Eq.(3) but from the most general
solutions which
---as we have already seen--- are {\em sums} of solutions (3) over
the various values $b_m$ of $b$, then a Lorentz transformation
will lead us to {\em a sum} of monochromatic beams: actually, of
harmonics (rather than to a single monochromatic beam). \ In
particular, if one wants to obtain a sum of harmonic beams, one
has to apply a LT to more general subluminal pulses.

\

\begin{figure}[!h]
\begin{center}
 \scalebox{.8}{\includegraphics{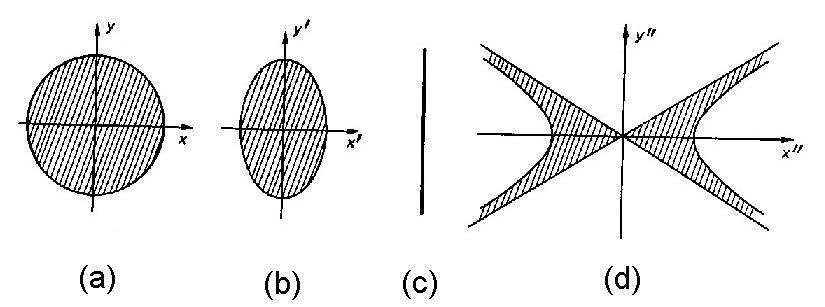}}
\end{center}
\caption{Let us consider an object that is intrinsically
spherical, i.e., that is a sphere in its rest-frame (Panel (a)).
After a generic subluminal LT along $x$, i.e., under a subluminal
$x$-boost, it is predicted by Special Relativity (SR) to appear as
ellipsoidal due to
Lorentz contraction (Panel (b)).  After a superluminal
$x$-boost\cite{Review,JWeber,NCim} (namely, when this object
moves\cite{RMDartora} with superluminal speed $V$), it is
predicted by SR ---in its non-restricted version (ER)--- to
appear\cite{Barut} as in Panel (d), i.e., as occupying the
cylindrically symmetric region bounded by a two-sheeted rotation
hyperboloid and an indefinite double cone. The whole structure,
according to ER, is expected to move {\em rigidly\/} and, of course, with
the speed $V$, the cotangent square of the cone semi-angle being
$(V/c)^2 - 1$. \ Panel (c) refers to the limiting case when the
boost-speed tends to $c$, either from the left or from the right
(for simplicity, a space axis is skipped). It is remarkable that
the shape of the localized (subluminal and superluminal) pulses,
solutions to the wave equations, appears to follow the same
behavior; this can have a role for a better comprehension even of
de Broglie and Schroedinger wave-mechanics. \ The present figure
is taken from refs.\cite{Barut,Review}. \ See also the next Figure.}
\label{fig5}
\end{figure}

\begin{figure}[!h]
\begin{center}
 \scalebox{.32}{\includegraphics{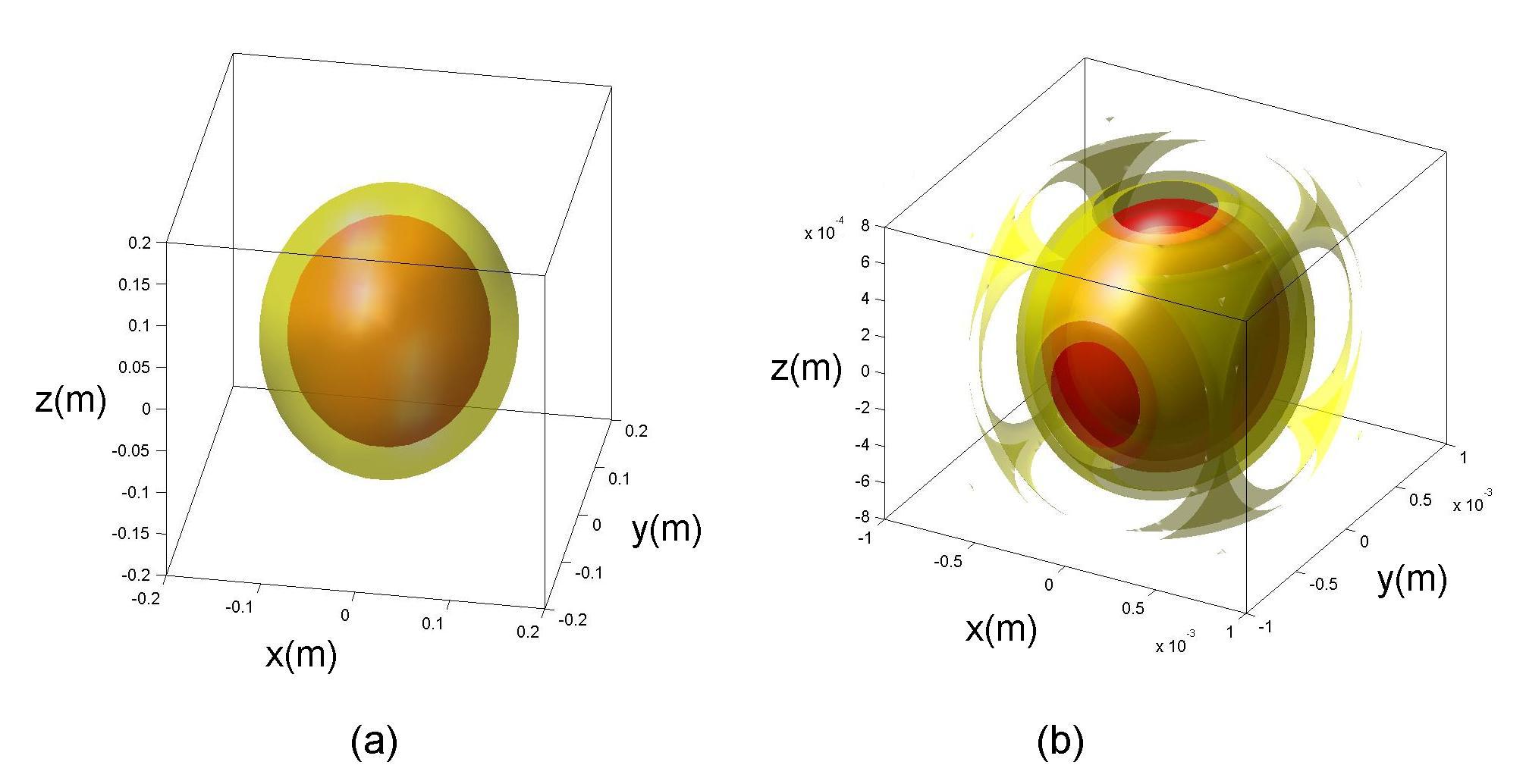}}
\end{center}
\caption{In the previous Figure we have seen how SR, in its non-restricted
version (ER), predicted\cite{Barut,Review} that, while the simplest
subluminal object is obviously a shere
(or, in the limit, a space point), the simplest superluminal object is
on the contrary an X-shaped pulse (or, in the limit, a double cone). \ The
circumstance that the Localized Solutions to the wave equations do follow the
same pattern is rather interesting, and is expected to be useful ---in
the case, e.g., of elementary particles and quantum physics--- for a deeper
comprehension of de Broglie's and Schroedinger's wave mechanics. With regard
to the fact that the simplest subluminal LWs, solutions to the wave equations,
are ``ball-like", let us depict by these figures, in the ordinary 3D
space, the general shape of the Mackinnon's solutions as expressed by
Eq.(10), numerically evaluated for $v<<c$.  In figures (a) and (b) we
graphically represent the field iso-intensity surfaces, which in the
considered case result to be (as expected) just spherical.}
\label{fig6}
\end{figure}

\h Let us add that {\em also} the various superluminal localized
pulses get transformed\cite{Saari2004} one into the other by the
mere application of ordinary LTs; while it may be expected that
the subluminal and the superluminal LWs are to be linked (apart
from some known technical difficulties, that require a particular
caution) by the superluminal Lorentz "transformations" expounded
long ago, e.g., in refs.\cite{Review,JWeber,NCim,Barut} and refs.
therein.\footnote{One should pay attention that, as we were
saying, the topic of superluminal LTs is a
delicate\cite{Review,JWeber,Ncim,Barut} one, at the extent that
the majority of the recent attempts to re-address this question
and its applications seem to be defective (sometimes they do not
even keep the necessary covariance of the wave equation itself).}
\ Let us recall once more that, in the years 1980-82, special
relativity, in its non-restricted version, predicted that, while
the simplest subluminal object is obviously a sphere (or, in the
limit, a space point), the simplest superluminal object is on the
contrary an X-shaped pulse (or, in the limit, a double cone): cf.
Fig.\ref{fig5}, taken from refs.\cite{Barut,Review}. \ The
circumstance that the pattern of the localized solutions to the
wave equations does meet this prediction is rather interesting,
and is expected to be useful ---in the case, e.g., of elementary
particles and quantum physics--- for a deeper comprehension of de
Broglie's and Schroedinger's wave mechanics. With regard to the
fact that the simplest subluminal LWs, solutions to the wave
equation, are ``ball-like", let us depict by Figs.\ref{fig6}, in
the ordinary 3D space, the general shape of the Mackinnon's
solutions as expressed by Eq.(10) for $v<<c$: In such figures we
graphically represent the field iso-intensity surfaces, which in
the considered case result to be (as expected) just spherical.

\

\h We have also seen, among the others, that, even if our first method
(Section 2) cannot yield {\em directly} zero-speed envelopes, such envelopes
``at rest", in Eq.(31), can be however obtained by applying a $v$-speed LT
to eq.(16). In this way, one starts from many frequencies [Eq.(16)] and
ends up with one frequency only [Eq.(31)], since $b$ gets transformed into
{\em the} frequency of the monochromatic beam.

\

\

\section{Non-axially symmetric solutions: The case of higher-order
Bessel beams}

\h Let us stress that till now we have paid attention to exact
solutions representing axially-symmetric (subluminal) pulses only:
that is to say, to pulses obtained by suitable superpositions of
zero-order Bessel beams.

\h It is however interesting to look also for analytic solutions
representing {\em non}-axially symmetric subluminal pulses, which
can be constructed in terms of superpositions of $\nu$-order
Bessel beams, with $\nu$ a positive integer ($\nu>0$). \ This can be
attempted both in the case of Sect.2 (first method), and in
the case of Sect.3 (second method).

\h For brevity's sake, let us take only the first methos (Sect.2)
into consideration. \ One is immediately confronted with the difficulty
that {\em no} exact solution is known for the integral in Eq.(8)
when $J_0(.)$ is replaced with $J_\nu (.)$.

\h One can overcome this difficulty by following a simple method,
which will allow us to obtain ``higher-order" subluminal waves in
terms of the axially-symmetric ones. \ Indeed, it is well-known that,
if $\Psi(x,y,z,t)$ is an exact solution to the ordinary wave equation, then \
$\partial^n \Psi/ \partial x^n$ \ and \ $\partial^n \Psi/ \partial y^n$ \
are also exact solutions.\footnote{Let us mention that even \
$\partial^n \Psi/ \partial z^n$ \ and \ $\partial^n \Psi/ \partial t^n$
will be exact solutions.} \ One should notice that, on the contrary, when
working in
cylindrical co-ordinates, if $\Psi(\rho,\phi,z,t)$ is a solution to the
wave equation, $\partial \Psi/ \partial \rho$ \ and \ $\partial \Psi/
\partial \phi$ are {\em not} solutions, in general. \ Nevertheless, it
is not difficult at all to reach the noticeable conclusion that,
once \ $\Psi(\rho,\phi,z,t)$ \ is a solution, then also

\bb \overline{\Psi}(\rho,\phi,z,t) \ug e^{i\phi}\left(
\frac{\partial \Psi}{\partial \rho} +
\frac{i}{\rho}\frac{\partial\Psi}{\partial \phi}\right) \label{ho}
\ee  

\

is an exact solution. For example, for an axially-symmetric solution
of the type $\Psi =
J_0(k_{\rho}\rho)\,\exp[ik_z]\,\exp[-i\om t]$, equation
(\ref{ho}) yields $\overline{\Psi} = - k_{\rho}
\,J_1(k_{\rho}\rho)\,\exp[i\phi]\,\exp[ik_z]\,\exp[-i\om t]$,
which is actually another analytic solution.

\h In other words, it is enough to start for simplicity from a
zero-order Bessel beam, and to apply Eq.(\ref{ho}), successively,
$\nu$ times, in order to get as a new solution \ $\overline{\Psi}
= (-k_{\rho})^\nu \, J_\nu(k_{\rho}\rho)\,\exp[i \nu
\phi]\,\exp[ik_z]\,\exp[-i\om t]$, which is a $\nu$-order Bessel
beam. \

\h In such a way, when applying $\nu$ times Eq.(\ref{ho}) to the
(axially-symmetric)
subluminal solution $\Psi(\rho,z,t)$ in Eqs.(16,15,14) \
[obtained from Eq.(3) with spectral function $S(\om)$], we are
able to get the subluminal non-axially symmetric pulses \
$\Psi_{\nu}(\rho,\phi,z,t)$ \ as new analytic solutions,
consisting as expected in superpositions of $\nu$-order Bessel
beams:

\bb  \Psi_{n}(\rho,\phi,z,t) \ug \int_{\om_-}^{\om_+} \drm \om \;
S'(\om)\;J_\nu(k_{\rho}\rho) \; e^{i \nu \phi} \; e^{ik_z z} \; e^{-i \om t} \; ,
\label{supho} \ee   

with $k_{\rho} (\om)$ given by Eq.(4), and quantities $S'(\om) =
(-k_{\rho}(\om))^\nu S(\om)$ being the spectra of the new pulses.
\ If $S(\om)$ is centered at a certain carrier frequency (it is a
gaussian spectrum, for instance), then $S'(\om)$ too will
approximately be of the same type.

\h Now, if we wish the new solution $\Psi_{\nu}(\rho,\phi,z,t)$ to
possess a pre-defined spectrum $S'(\om) = F(\om)$, we can first
take Eq.(3) and put \ $S(\om) = F(\om) / (-k_{\rho}(\om))^\nu$ \
in its solution (16), and afterwards apply to it, $\nu$ times, the
operator \ $U \equiv \exp[i\phi] \; [\partial / \partial\rho + (i
/ \rho) \partial /
\partial\phi)] \, $: \ As a result, we will obtain the desired
pulse, $\Psi_{\nu}(\rho,\phi,z,t)$, endowed with $S'(\om) =
F(\om)$.

\

{\bf An example:}

\

On starting from the subluminal axially-symmetric
pulse $\Psi(\rho,z,t)$, given by Eq.(16) with the {\em gaussian}
spectrum (17), we can get the subluminal, non-axially symmetric,  exact
solution $\Psi_{1}(\rho,\phi,z,t)$ by simply calculating

\bb \Psi_{1}(\rho,\phi,z,t) \ug {{\partial \Psi} \over
{\partial \rho}} \ \; e^{i\phi} \; , \label{ho1} \ee  

which actually yields the ``first-order" pulse $\Psi_{1}(\rho,\phi,z,t)$,
which can be more compactly written in the form:

\bb \Psi_{1}(\rho,\phi,\eta,\zeta) \ug 2 \, {b \over c} \, v \,
\gamma^2 \; {\exp{\left[ i{b \over c} \, \beta \, \gamma^2 \; \eta
\right]}} \sum_{n=-\infty}^{\infty} \; A_n \; \exp{[in{\pi \over
\beta}]} \ \; \psi_{1n}
\label{eq.(52)} \ee   

with

\bb \psi_{1n}(\rho,\phi,\eta,\zeta) \equiv \dis{ {b^2 \over c^2}
\, \gamma^2 \rho  \ \; Z^{-3} \ [ Z \, \cos Z - \sin Z ] \ \;
e^{i\phi} }
 \; , \label{eq.(53)} \ee  

where

\bb Z \equiv \dis{ \sqrt{{b^2 \over c^2} \, \gamma^2 \rho^2 +
\left({b \over c} \, \gamma^2 \zeta + n \pi  \right)^2 }} \; .
\label{eq.(54)}
\ee  

\begin{figure}[!h]
\begin{center}
 \scalebox{.25}{\includegraphics{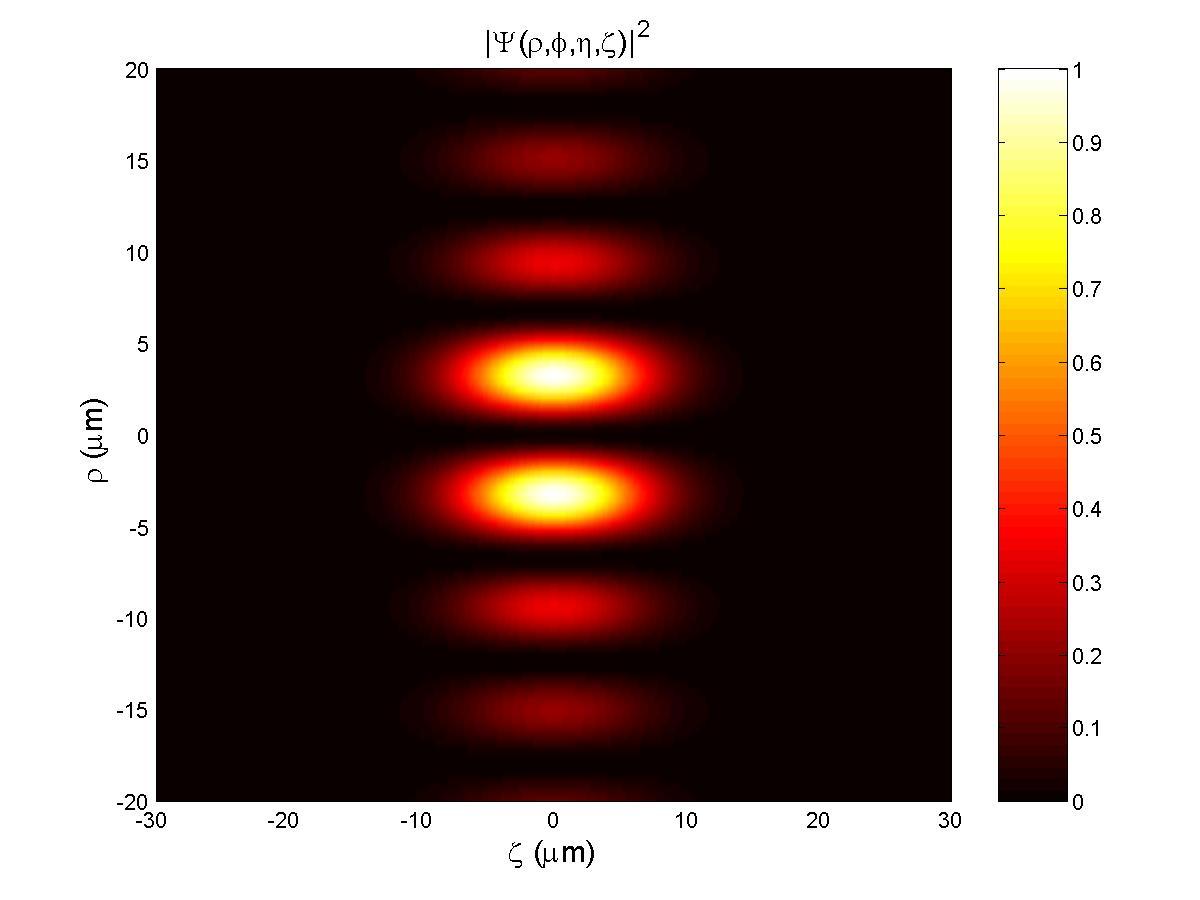}}
\end{center}
\caption{Orthogonal projection of the field intensity
corresponding to the higher order subluminal {\em pulse} represented by
the exact solution Eq.(51), quantity $\Psi$ being given by Eq.(16) with
the gaussian spectrum (17). The pulse intensity happens to have this
time a ``donut"-like shape.} \label{fig7}
\end{figure}

\h This exact solution, let us repeat, corresponds to
superposition (\ref{supho}), \ with \ $S'(\om) = k_{\rho}(\om)
S(\om)$, \ quantity $S(\om)$ being given by Eq.(\ref{eq.(17)}). \
It is represented in Figure \ref{fig7}. The {\em pulse} intensity has
a ``donut-like" shape.

\

\

\section{Conclusions}

\h Like in the well-known superluminal case, the (more orthodox,
in a  sense) {\em subluminal} Localized Waves can be obtained by
suitable  superpositions of Bessel beams. They had been till now
almost neglected, however, for  the mathematical difficulties met
in finding out analytic expressions for  them, difficulties
associated with the fact that the superposition integral
 runs over a finite interval. We have re-addressed in this paper the
question  of such subluminal LWs, obtaining by contrast, and in a
simple way, non-diffracting subluminal pulses as exact analytic
solutions to the wave equations. Such new ideal subluminal
solutions ---which propagate without distortion in any homogeneous
linear medium--- have been herein obtained for arbitrarily chosen
frequencies and bandwidths, avoiding in particular any recourse
 to the non-causal components so frequently plaguing the previously known
 localized waves.

\h Indeed, only one closed-form subluminal LW solution, $\psi_{\rm
cf}$, to the wave equations was known\cite{Mackinnon}: the one
obtained  by choosing in the relevant integration a constant
weight-function  $S(\om)$; while all other solutions had been got
in the past only via numerical simulations. By contrast, we have
shown above, for instance, that a subluminal LW can be obtained in
closed form by adopting any spectra  $S(\om)$ that be expansions
in terms of $\psi_{\rm cf}$. \ In fact, the initial disadvantage
of the subluminal case, that of being associated with a a limited
bandwidth, has been turned into an advantage, since in the case of
``truncated" integrals ---at variance with the superluminal
case--- the spectrum $S(\om)$ can be expanded in a Fourier series.

\h More in general, it has been shown in what precedes how can one
arrive at exact solutions both by integration over the Bessel
beams' angular frequency $\om$, and by integration over their
longitudinal wavenumber $k_z$. Both methods are expounded in this
paper. The first one appears to be powerful enough; we have
studied the second method as well, however, since it allows
dealing even with the limiting case of
zero-speed solutions (thus furnishing a new way, in terms of
continuous spectra, for obtaining the so-called ``Frozen Waves",
so promising also from the point of view of applications). \  We have
briefly treated the case, moreover, of non axially-symmetric
solutions, namely, of higher order Bessel beams.

\h At last, attention has been always paid to the role of Special
 Relativity, and to the fact that the localized waves are transformed one
 into the other by suitable Lorentz Transformations. Moreover, our
results seem to show that in the subluminal case the {\em
simplest} LW solutions are (for $v<<c$) ``ball"-like, as expected
since long\cite{Barut} on the mere basis of special
relativity\cite{Review}.  Actually, already in the years
 1980-82 it had been predicted that, if the simplest subluminal
 object is a sphere (or, in the limit, a space point), then the simplest
 superluminal object is an X-shaped pulse (or, in the limit, a double-cone);
 and viceversa: Cf. Fig.5, taken from ref.\cite{Barut}.  It is
rather interesting that the same pattern appears to be followed by the
localized solutions of the {\em wave equations}.  For the subluminal case,
see. e.g., Figs.6.

\h The localized pulses, endowed with a finite energy, or merely
 truncated, will be constructed in another paper.

\h In the present work we have fixed our attention especially on
electromagnetism and optics: but results of the same kind are
valid whenever an essential role is played by a wave-equation
(like in acoustics, seismology, geophysics, gravitation,
elementary particle physics, etc.).

\

\section{Acknowledgements}

The authors are indebted to H.E. Hern\'andez-Figueroa for his
continuous collaboration, and thank I.M.Besieris, A.Castoldi, C.Conti,
J-y.Lu, G.C.Maccarini, R.Riva, P.Saari, A.M.Shaarawi, M.Tygel for useful
discussions or kind interest.

\end{document}